\documentclass[twocolumn,showpacs,preprintnumbers,amsmath,amssymb]{revtex4}

\usepackage{graphicx}
\usepackage{dcolumn}
\usepackage{bm}
\usepackage{amsmath}

\begin{document}


\title{Point-defect haloing in curved nematic films}

\author{Isaku Hasegawa}
\email{isaku@eng.hokudai.ac.jp}
\affiliation{Department of Applied Physics, Graduate School of 
Engineering, Hokkaido University, Sapporo 060-8628, Japan}

\author{Hiroyuki Shima}
\affiliation{Department of Applied Physics, Graduate School of 
Engineering, Hokkaido University, Sapporo 060-8628, Japan}
\affiliation{Department of Applied Mathematics 3, LaC$\grave{a}$N, 
Universitat Polit$\grave{e}$cnica de Catalunya (UPC), Barcelona 
08034, Spain}

\date{\today}

\begin{abstract}
We investigate the correlation between the point disclination energies and the 
surface curvature modulation of nematic liquid crystal membranes with a 
Gaussian bump geometry.
Due to the correlation, disclinations feel an attractive force that confines 
them to an annulus region, resulting in a halo distribution around the 
top of the bump.
The halo formation is a direct consequence of the nonzero Gaussian curvature 
of the bump that affects preferable configurations of liquid crystal molecules 
around the disclination core.
\end{abstract}

\maketitle
\section{Introduction}
It is generally assumed that topological defects in liquid crystal membranes 
are excluded in the ground state because they increase the elastic free energy 
of the system to more than that in the defect-free equilibrium state.
The situation is quite different when the liquid crystal phases are embedded 
in a curved surface.
In a curved membrane with nonzero Gaussian curvature, the appearance of 
topological defects can be favorable or even inevitable because of 
geometrical and/or topological constraints.
A typical example is a disclination-containing equilibrium state in a nematic 
liquid crystal that is constrained in a spherical surface 
\cite{lubensky}.
The defect structure in such a nematic spherical shell has recently been 
reconsidered both theoretically \cite{nelson,vitelli1,skacej,shin} and 
experimentally \cite{fernandeznieves,lopezleon}, and these investigations 
revealed richer physical properties than those that have initially been 
suggested.
In addition to the nematic cases, the smectic phases of liquid crystal films 
\cite{xing, gomez} and columnar phases of a block copolymer assembly 
\cite{santangelo} have also 
exhibited an interplay between disclination structures and the 
geometry/topology of the constraint surface.
In particular, smectic ordering on a sinusoidal substrate has indicated a 
direct coupling between the substrate curvature and the preferential 
locations of disclinations in equilibrium states \cite{gomez}.
The coupling is purely a geometric curvature effect in the sense that the 
system involves no topological requirement for the existence of disclination.
The geometric curvature also strongly influences the orientational order of 
the spin lattice models defined on curved surfaces 
\cite{shima1,shima2,dandoloff,isaku,baek1,carvalhosantos,baek2,sakaniwa}.

Many intriguing results have been obtained with regard to the global 
lowest-energy state on a curved surface with a minimum number of 
disclinations, which should be relevant to liquid crystal phases at very low 
temperatures.
In actual systems, however, a slight physical (and chemical) disturbance 
causes excess disclinations \cite{degennes,mermin,romanov,kleman}.
These disclinations can travel over a surface via thermal excitation, and they 
flow almost freely if the disclination density is sufficiently low to ignore 
their interaction.
The only possible force that restricts the free motion of the excess 
disclinations arises from a coupling between the surface curvature and the 
disclination energy.
The substrate curvature breaks the translational symmetry of the molecular 
configuration in the vicinity of the disclination core, which may cause a 
shift in the disclination energy.
As a result, the energetically preferential positions of excess disclinations 
are believed to be tunable by imposing an appropriate curvature modulation.

In this article, we study the energies of disclinations in a nematic thin 
film assigned on a Gaussian bump.
It is found that disclinations with a $+1$ charge prefer positioning at 
inflection points of the bump, implying the formation of a \lq\lq halo,\rq\rq 
i.e., an annulus distribution of excess disclinations around the top of the 
bump.
In addition, disclinations with a $-1$ charge are found to exhibit double 
concentric halo structures, provided the bump height is relatively large.
These halo formations are consequences of the non-trivial coupling between 
the disclination energies and the surface curvature modulation.

\begin{figure}[b]
\begin{center}
\includegraphics[width=250pt]{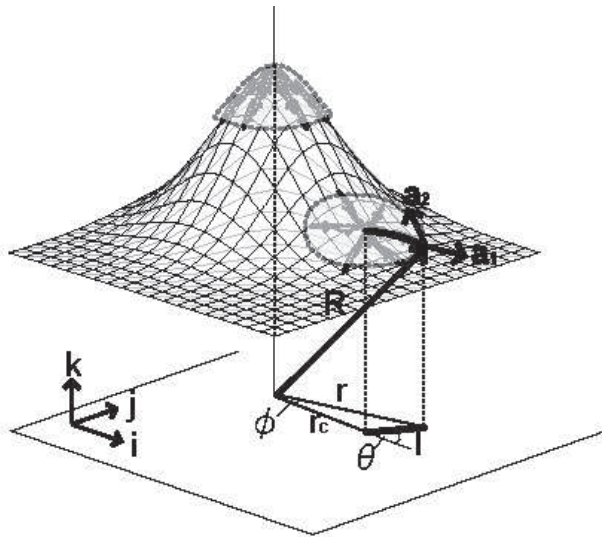}
\end{center}
\caption{\label{fig:geometry} Schematic of a Gaussian bump and mathematical 
notations used in the paper. 
The thick arc depicted on the bump is a projection of the thick line segment 
on the underlying flat plane.
}
\end{figure}

\section{Model}\label{sec:elenvefi}
We consider nematic liquid crystals confined in a thin curved membrane with 
thickness $h$, in which the molecules align parallel to the tangential 
directions independently of the depth in the membrane.
In line with the continuum theory \cite{young,degennes}, the elastic free 
energy of the molecular ensemble within the volume $Ah$ is given by
\begin{eqnarray}
F_d = h \int_A dA \left\{ \frac{K_1}{2} ({\bm D} \cdot {\bm n})^2
+ \frac{K_3}{2} ({\bm D} \times {\bm n})^2 \right\}
\mbox{,}
\label{eq:energy}
\end{eqnarray}
where the unit vector $\bm n$ (called the director field) describes the 
local molecular alignment.
$\bm D$ is a vector operator whose components $D_{\mu}$ are the covariant
derivative \cite{shima3} on the curved surface.
$K_1$ and $K_3$ are the elastic constants associated with splay and bend 
distortions of the molecular configuration, respectively \cite{stewart}.

Suppose that the director lies within a thin curved shell having a Gaussian 
bump geometry.
Points on the bump are represented by a three-dimensional vector 
$\bm R(r,\phi)$ given by
\begin{equation}
{\bm R(r,\phi)} = r \cos \phi \bm{i} + r \sin \phi \bm{j}
                     + v \exp \biggl(-\frac{r^2}{2 r_0^2} \biggr) \bm{k},
                      \label{eq:points}
\end{equation}
where $\bm{i}$, $\bm{j}$, and $\bm{k}$ are basis vectors of the 
Cartesian coordinate system, and $r$ and $\phi$ are the polar coordinates 
assigned on the underlying flat plane (see Fig.\ref{fig:geometry}).
The geometric deviation of a bump from a flat plane is characterized by a 
dimensionless parameter $\alpha \equiv v/r_0$ that represents the ratio of the 
height to the width of the bump.
Note that the constant $r_0$ in the last term in Eq.(\ref{eq:points}) 
determines the points of inflection of the Gaussian bump, i.e., across which 
the sign of $(\partial^2 {\bm R} / {\partial r}^2) \cdot {\bm k}$ changes.
We demonstrate below that $r_0$ plays a prominent role in describing the 
correlation between the disclination energy and the substrate curvature.

Let us locate a disclination core on a point 
${\bm R} = {\bm R}(r_c,\phi_c)$, where $\phi_c$ is an arbitrary 
constant without loss of generality.
The disclination size is sufficiently smaller than the bump width, and the 
perimeter of the disclination is defined by the locus of points whose 
distances from the core are equal in the sense of geodesic distance; 
therefore, the perimeter slightly deviates from a circular shape when it is 
placed on a curved surface, whereas it is an exact circle when placed on a 
flat plane.

\begin{figure*}[t]
\begin{center}
\includegraphics[width=400pt]{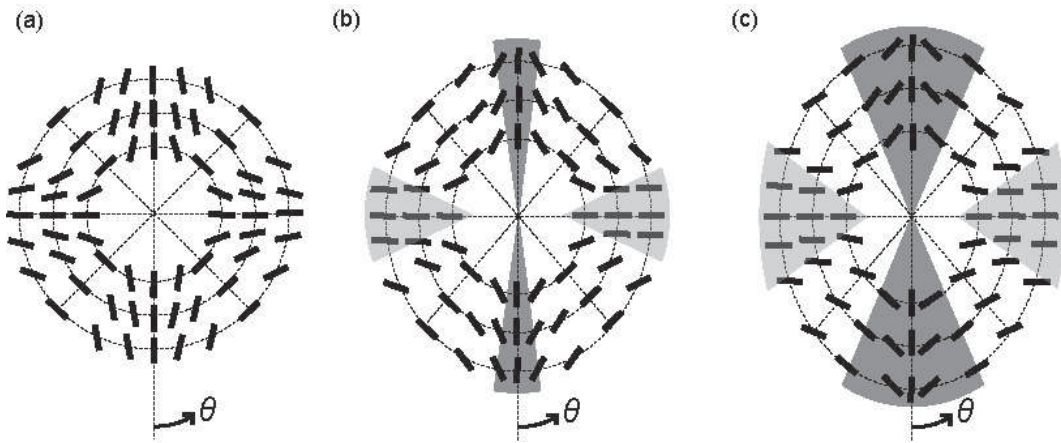}
\end{center}
\caption{\label{fig:sketchv} Configurations of $\bm n$ in a disclination 
with a charge of $- 1$.
The core positions $r_c$ are $r_c = 0$ in (a), $0 < r_c < r_0$ in (b), and 
$r_c = r_0$ in (c).
In the light- and dark-shadowed region, the splay deformation is suppressed 
and enhanced, respectively.}
\label{fig:ConfDiM}
\end{figure*}

For later use, we introduce a new polar coordinate system $(l,\theta)$ on the 
flat plane such that the coordinate origin (i.e., $l=0$) locates at the point 
of interaction of the flat plane and the vertical line penetrating the 
disclination core (see Fig.\ref{fig:geometry}).
Then, the basis vectors
${\bm a}_1 \equiv \partial {\bm R}/\partial l$ and 
${\bm a}_2 \equiv \partial {\bm R}/\partial \theta$ span the 
curved surface and allow us to write ${\bm n} = n^{i} {\bm a}_i$ 
($i=1,2$), where $n^i$ and ${\bm a}_i$ depend on $l$ and $\theta$ (See 
Appendix \ref{app:nnonG}).
The director orientation around the core is given by the angle $\psi$ 
between $\bm{n}$ and ${\bm a}_1$.
It is natural to assume that
\begin{eqnarray}
 \psi &=& (s-1) \theta + \psi_0,
 \label{eq:psi}
\end{eqnarray}
where $s$ denotes the disclination charge that quantifies how many times 
$\bm n$ rotates about the core.
When $s = +1$, for instance, all ${\bm n}$ contained within the disclination 
area $A$ make the same angle $\psi_0$ with respect to ${\bm a}_1$ (see 
Fig.\ref{fig:ConfDiP}(a)); on the other hand, when $s=-1$, the angle increases 
proportionally to $\theta$ (see Fig.\ref{fig:ConfDiM}(a)).

In the calculation, we consider disclinations with two charges, $s = \pm 1$.
The disclination area $A$ and the membrane thickness $h$ are 
set to be $A = 400 \pi \  \rm{\mu m^2}$ and $h = 4 \  \rm{\mu m}$ by 
referring to the material constants of actual liquid crystal membranes 
\cite{shah}.
The values of the elastic constants $K_1$ and $K_3$ are those of 
4-methoxybenzylidene-4$^{\prime}$-butylaniline (MBBA), p-azoxyanisole (PAA), 
and 4-pentyl-4$^{\prime}$-cyanobiphenyl (5CB) presented in 
Table \ref{tab:values} \cite{stewart}.
The constant $\psi_0$ in Eq.(\ref{eq:psi}) is fixed to be zero without loss of 
generality, bearing in mind the fact that other choices of $\psi_0$ lead to no 
essential change in the conclusion of this work.

\begin{table}[b]
\caption{\label{tab:values} Elastic constants of MBBA, PAA, and 5CB reported 
in Ref.\cite{stewart}}
\begin{ruledtabular}
\begin{tabular}{cccc}
Elastic constants & MBBA & PAA & 5CB \\
(pN) & ($25^\circ$C) & ($122^\circ$C) & ($26^\circ$C)\\
\hline
$K_1$ & 6 & 6.9 & 6.2 \\
$K_3$ & 7.5 & 11.9 & 8.2 \\
\end{tabular}
\end{ruledtabular}
\end{table}

\begin{figure*}[t]
\hspace{-20pt}
\begin{minipage}[t]{.45\linewidth}
\begin{center}
\includegraphics[width=210pt]{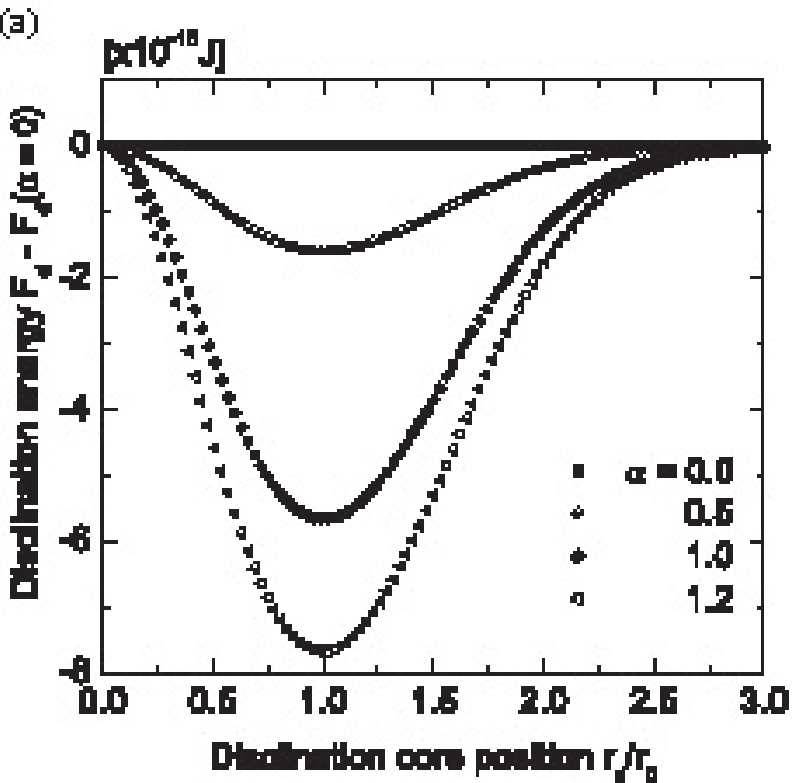}
\end{center}
\end{minipage}
\hspace{10pt}
\begin{minipage}[t]{.45\linewidth}
\begin{center}
\includegraphics[width=210pt]{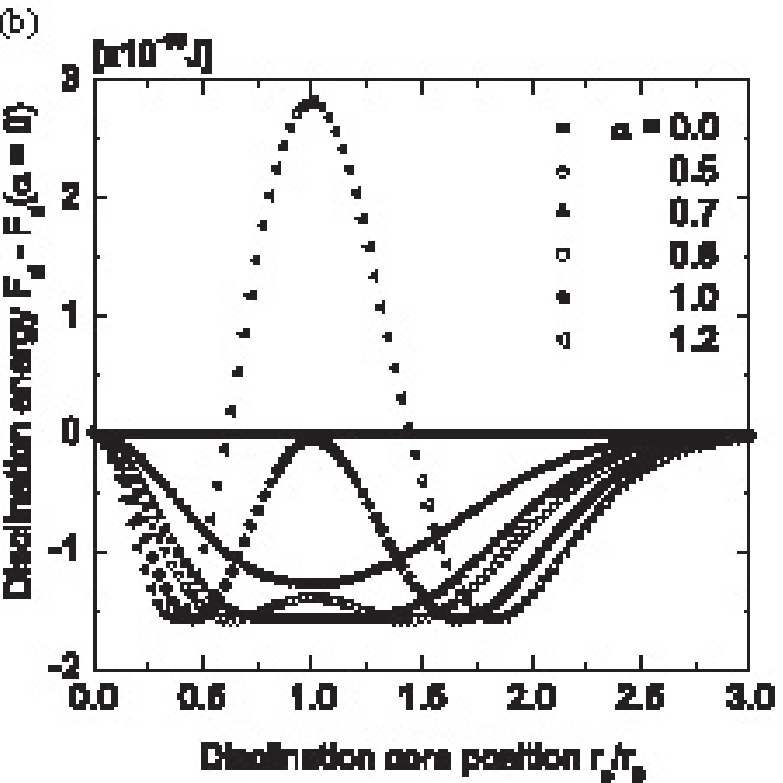}
\end{center}
\end{minipage}
\caption{\label{fig:energyplot} Disclination energy in deformed MBBA films as 
a function of the core position $r_c$. The parameter $\alpha$ characterizing 
the bump height ranges from $0$ to $1.2$ (as indicated).
(a) Energy of a positively charged disclination ($s=+1$). 
The magnitude of a downward peak at $r_c = r_0$ monotonically increases with 
$\alpha$.
(b) Energy of a negatively charged disclination ($s=-1$).
Transition from a single-well to a double-well structure is observed at 
$\alpha = 0.7$.}
\end{figure*}

\section{Result}
Figure \ref{fig:energyplot} shows plots of the disclination energy $F_d$ of 
MBBA as a function of the core position $r_c$.
The vertical axis shows the difference in $F_d$ between the bump and the flat 
plane, i.e., $F_d(\alpha \ne 0)-F_d(\alpha=0)$.
The disclination charge $s$ is $s = +1$ in (a) and $-1$ in (b), and the 
parameter $\alpha$ ranges from $\alpha = 0$ to $1.2$ in both (a) and 
(b).
For $s = +1$, $F_d$ takes the minimum at $r_c = r_0$ for all values of 
$\alpha$, indicating that a disclination is stable at the inflection 
point of the bump $r = r_0$ independently of the bump height.
On the other hand, the plot for $s = -1$ exhibits a transition from a 
single downward peak structure ($\alpha < 0.7$) into a double-well structure 
($\alpha > 0.7$) such that $r_0$ becomes unstable.
The contrasting behaviors of $F_d$ for $s = +1$ and $-1$ are attributed to 
surface curvature effects on the director configuration close to the core, as 
will be shown in Sec.\ref{sec:discu}.
For PAA and 5CB, we have obtained similar profiles of $F_d$ exhibiting a 
downward peak at $r_c = r_0$ for $s = +1$ and a double-well structure for 
$s = -1$, whereas the peak heights differ slightly from those shown in 
Fig.\ref{fig:energyplot} because of quantitative differences in 
$K_1$ and $K_3$.
We have also confirmed that the peak positions (both downward and upward) 
are independent of our choice of $\psi_0$ defined by Eq.(\ref{eq:psi}).

\begin{figure}[b]
\begin{center}
\includegraphics[width=220pt]{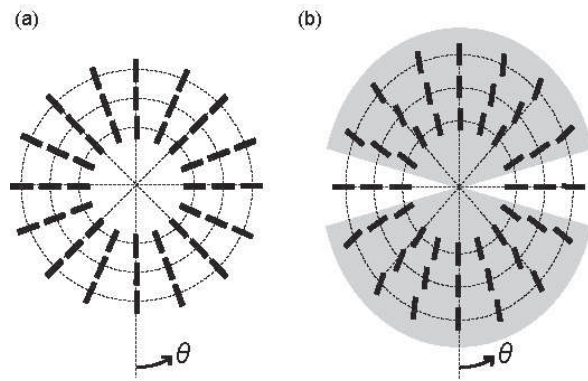}
\end{center}
\caption{Configurations of the director $\bm n$ 
around the core of a disclination with a charge of $+ 1$.
With a shift in the core position from $r_c = 0$ (a) to $r_c = r_0$ (b), the 
radial symmetry of the director configuration breaks down such that the splay 
deformation in the shadowed region in (b) is suppressed.}
\label{fig:ConfDiP}
\end{figure}

We emphasize that the energy scales of the potential wells shown in 
Figs.\ref{fig:ConfDiP}(a) and \ref{fig:ConfDiP}(b) exceed the thermal energy 
scale at ambient temperature.
For instance, the depth of the single well for $s=+1$ and $\alpha=1.0$ given 
in Fig.\ref{fig:ConfDiP}(a) is approximately $5.7 \times 10^{- 18} \rm{J}$; 
this is of the order of $10^{3} \  k_B T$ at $T=25^\circ\rm{C}$.
This fact suggests that the potential well at $r=r_0$ overwhelms the thermal 
excitation that causes a random movement of disclinations over the surface.
As a result, an ensemble of trapped disclinations forms a halo with radius 
$r_c$ around the top of the Gaussian bump, provided that the disclination 
density is sufficiently low to ignore their repulsive interaction.
The same scenario applies to disclinations with a charge of $-1$, implying the 
formation of doubly consentric halos observed for $\alpha>1.0$.

\begin{figure}[b]
\hspace{-15pt}
\begin{minipage}[t]{.45\linewidth}
\begin{center}
\includegraphics[width=120pt]{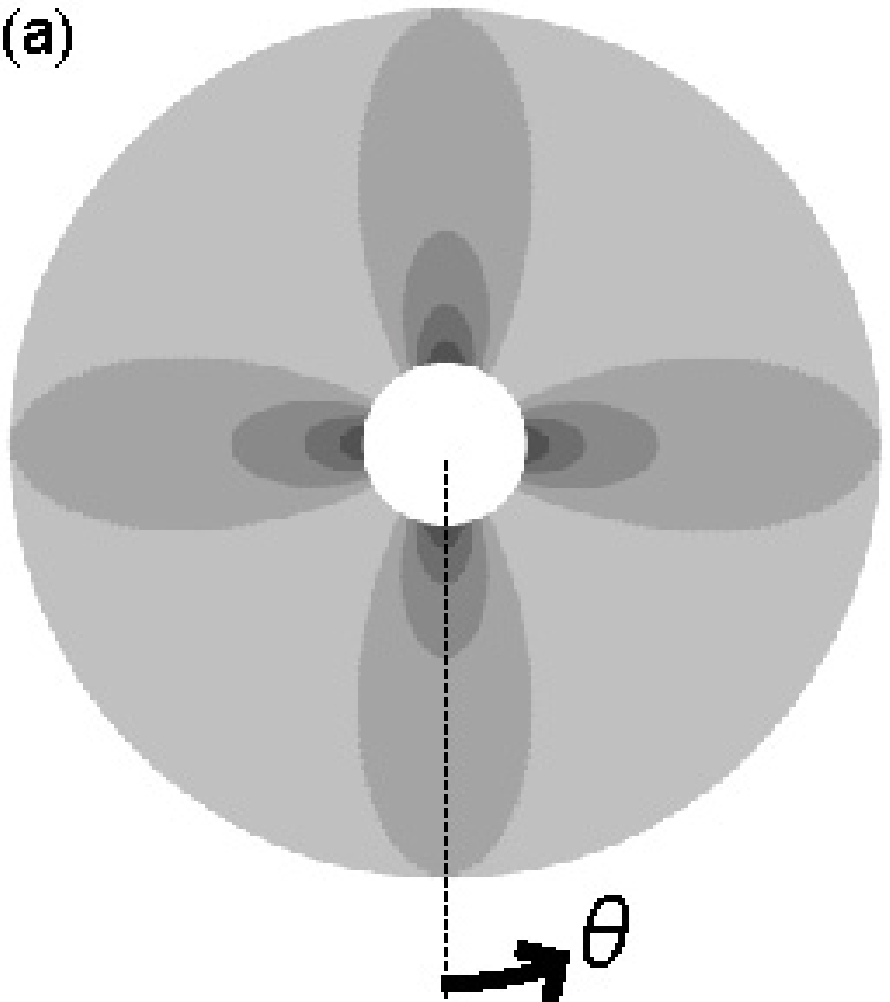}
\end{center}
\end{minipage}
\hspace{10pt}
\begin{minipage}[t]{.45\linewidth}
\begin{center}
\includegraphics[width=120pt]{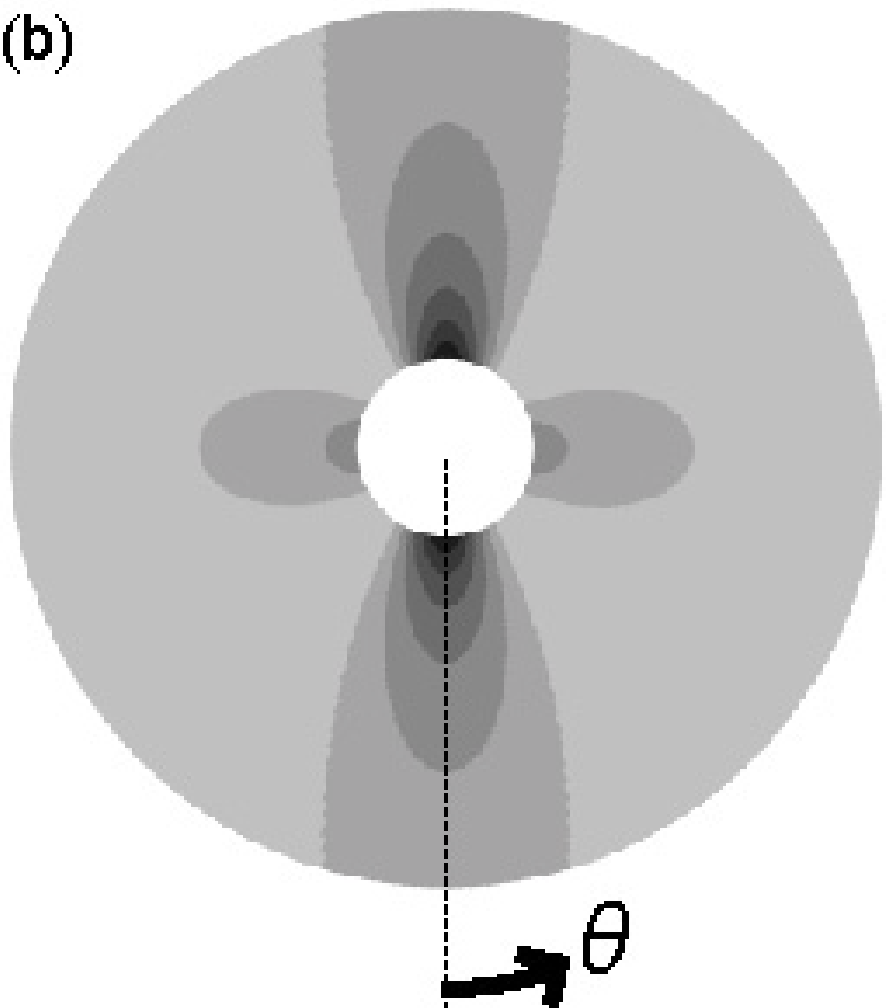}
\end{center}
\end{minipage}
\caption{\label{fig:Mcontour} Contour plots of the splay term 
$K_1({\bm D} \cdot {\bm n})^2$ of disclinations with $s = -1$ placed 
at $r_c = 0$ (a) and at $r_c = r_0$ (b).
Light (dark) gray indicates a small (large) value of the term at the position.
With an increase in $r_c$, the splay term decreases around the line segments 
of $\theta = \pi/2 \ \rm{and} \ 3\pi/2$ but 
increases around those of $\theta = 0 \ \rm{and} \ \pi$.}
\end{figure}

\section{Discussions}
\label{sec:discu}
The physical origin of a downward peak in the plot of $F_d$ for $s = +1$ (see 
Fig.\ref{fig:energyplot}(a)) is a curvature-induced shift in the director 
configuration around the disclination core.
Figures \ref{fig:ConfDiP}(a) and \ref{fig:ConfDiP}(b) illustrate how the 
configuration varies with a shift in the core position $r_c$ from $r_c = 0$ 
(a) to $r_c = r_0$ (b).
In Fig.\ref{fig:ConfDiP}(a), $\bm n$ aligns radially around the top of 
the bump, exhibiting rotational symmetry such that 
${\bm n} \parallel {\bm a}_1$ and 
${\bm n} \cdot {\bm a}_2 = 0$ at every point.
This symmetry breaks down for $r_c>0$.
By increasing $r_c$, the perimeter of the disclination distorts slightly 
and becomes elliptical, and then, the relationship 
${\bm n} \cdot {\bm a}_2 = 0$ no longer holds (see 
Fig.\ref{fig:ConfDiP}(b)), except along the four specific line segments 
($\theta = 0,\ \frac{\pi}{2},\ \pi,\ \frac{3 \pi}{2}$).
The most important observation in Fig.\ref{fig:ConfDiP}(b) is a suppression of 
the splay deformation in the shadowed region.
The area of the corresponding region is maximized at $r_c = r_0$, as a result 
of which a downward peak $F_d$ arises at $r_c = r_0$ at the expense of a 
slight splay enhancement within a limited region around 
$\theta = \pm \frac{\pi}{2}$.

Parallel arguments to the preceding paragraph account for the double-well 
structure of $F_d$ for $s=-1$ that is shown in 
Fig.\ref{fig:energyplot}(b).
Figures \ref{fig:sketchv}(a)--(c) show the variation of the ${\bm n}$ 
configuration under $s = -1$ and $\alpha > 0.7$, each figure corresponding to 
the core position: $r_c = 0$ (a), $0 < r_c < r_0$ (b), and $r_c = r_0$ (c).
The director configuration in (a) exhibits a four-fold rotational symmetry 
that breaks down as $r_c$ increases.
Symmetry breaking induces a suppression of the splay deformation in the 
light-shadowed regions shown in Fig.\ref{fig:sketchv}(b) and 
\ref{fig:sketchv}(c) in a manner similar to the case of $s=+1$.
In contrast, the splay deformation in the dark-shadowed regions is 
enhanced with an increase in $r_c$.
These two competing effects result in the double-well structure of $F_d$ for 
$s=-1$ cases.
That is, $F_d$ decreases with $r_c$ when the suppression effect exceeds the 
enhancement, and it increases in the opposite situation.
The degree of splay enhancement in the dark regions is maximized at $r_c=r_0$; 
this explains why $r_0$ is the most unstable point for $s=-1$ disclinations.

We have numerically confirmed that the splay term 
$K_1({\bm D} \cdot {\bm n})^2$ in Eq.(\ref{eq:energy}) plays a dominant 
role in describing the $r_c$- (and $\alpha$-) dependences of $F_d$.
Figures \ref{fig:Mcontour}(a)--(b) show contour plots of the splay term 
$K_1({\bm D} \cdot {\bm n})^2$ within the disclination area of $s=-1$.
We observe that with an increase in $r_c$, the local splay energy decreases at 
points near $\theta=\pi/2 \ \mbox{or} \ 3\pi/2$ but increases near 
$\theta=0 \ \mbox{or} \ \pi$; this is in agreement with our previous 
discussions.
A more detailed version of the study that covers the variations of the 
disclination size $A$ and the initial angle $\psi_0$ will be presented 
elsewhere.

\section{Summary}
In conclusion, we investigated the elastic energy of disclinations that arise 
in a curved nematic membrane with a Gaussian bump geometry.
When $\alpha<0.7$, the disclination energy $F_d$ exhibits a downward peak at 
$r=r_0$ independently of the sign of the disclination charge, suggesting a 
curvature-driven halo formation of disclinations around the top of the bump.
Furthermore, when $\alpha > 0.7$, $F_d$ for $s=-1$ exhibits a double-well 
structure in which the point $r=r_0$ becomes unstable in contrast to the case 
of $s=+1$; this implies doubly concentric halos peculiar to negatively charged 
disclinations.
In both cases of $s = \pm 1$, the well depth overcomes the thermal energy 
scale at room temperature, suggesting the experimental feasibility of our 
theoretical predictions.

\section*{Acknowledgements}
We would like to thank for K.~Yakubo and H. Orihara for the fruitful 
discussions.
IH is thankful for the financial support from the 21st Century COE Program 
\lq\lq Topological Science and Technology\rq\rq and support from the Japan 
Society for the Promotion of Science for Young scientists.
HS acknowledges M. Arroyo for assisting with the work carried out of UPC.
This work is supported by the Kazima foundation and a Grant-in-Aid for 
Scientific Research from the Japan Ministry of Education, Science, Sports and 
Culture.

\appendix
\section{Metric tensor 
$g_{\mu \nu} = {\bm a}_{\mu} \cdot {\bm a}_{\nu}$}
\label{app:eeaa}
In this Appendix, we derive the metric tensor 
$g_{\mu \nu} = {\bm a}_{\mu} \cdot {\bm a}_{\nu}$ in terms of the 
covariant basis vectors ${\bm a}_1 = \partial {\bm R}/\partial l$ and 
${\bm a}_2 = \partial {\bm R}/\partial \theta$ introduced 
in Sec. \ref{sec:elenvefi}.
We start with the basis vectors 
${\bm e}_1 = \partial {\bm R}(r,\phi)/\partial r$ and 
${\bm e}_2 = \partial {\bm R}(r,\phi)/\partial \phi$ associated with 
the polar coordinate system $(r,\phi)$ on the flat plane.
In terms of ${\bm e}_1$ and ${\bm e}_2$, the metric tensor 
${\tilde g}_{\mu \nu} = {\bm e}_{\mu} \cdot {\bm e}_{\nu}$ becomes
\begin{eqnarray}
{\tilde g}_{\mu \nu} = \left(
                  \begin{array}{cc}
                    f(r) & 0 \\
                    0 & r^2
                  \end{array}
                \right) \mbox{,}
\hspace{4pt}
f(r) = 1 + \frac{\alpha^2 r^2}{r_0^2} \exp \biggl(- \frac{r^2}{r_0^2} \biggr)
\mbox{.}
\end{eqnarray}
We use a coordinate transformation from $(r,\phi)$ to $(l,\theta)$ by 
shifting the coordinate origin from $r=0$ to $r=r_c$.
This is accomplished by multiplying both sides of ${\tilde g}_{\mu \nu}$ by 
the Jacobian ${\underline \beta}$ defined by
\begin{eqnarray}
{\underline \beta} = \left(
                        \begin{array}{cc}
           \partial r/\partial l & \partial \phi/\partial l \\
           \partial r/\partial \theta & \partial \phi/\partial \theta
                        \end{array}
                       \right) \mbox{,}
\end{eqnarray}
whose elements can be computed by using the following relationships:
 $r=\sqrt{r_c^2 + 2r_c l \cos \theta + l^2}$ and 
$\phi = \tan^{-1} \{l \sin \theta / (r_c + l \cos \theta) \}$.
Finally we obtain
\begin{eqnarray}
g_{\mu \nu} &=& {\bm a}_{\mu} \cdot {\bm a}_{\nu}
   = {\underline \beta} {\tilde g}_{\mu \nu} ({\underline \beta})^{-1} \cr \cr
  &=& \frac{1}{r^2} \left(
                 \begin{array}{cc}
   f(r)p^2 + q^2
     & l p q (1-f(r)) \\
       l p q (1-f(r))
& l^2 (f(r) q^2 + p^2)
                 \end{array}
                \right)
\label{eqapp:gmunu}
\mbox{,}
\end{eqnarray}
where $p=r_c \cos \theta + l$ and $q=r_c \sin \theta$.
In particular, when $\alpha=0$ (i.e., a flat plane), $f(r) \equiv 1$, 
and thus, we have 
\begin{equation}
g_{\mu \nu} = \left(
              \begin{array}{cc}
               1 & 0 \\
               0 & l^2
              \end{array}
              \right) \mbox{,}
\end{equation}
as expected.

\section{Contravariant components $n^i$}
\label{app:nnonG}
We present the derivation of the contravariant component $n^i$ of the 
vector ${\bm n} = n^i {\bm a}_i$ in terms of ${\bm a}_i$.
First, the relative angle between $\bm n$ and ${\bm a}_1$ is denoted as $\psi$.
Similarly, the angle between ${\bm a}_1$ and ${\bm a}_{2}$ is denoted
as $\Psi$.
It follows from Eq.(\ref{eqapp:gmunu}) that
\begin{eqnarray}
\cos{\Psi} &=&
 \frac{{\bm a}_1 \cdot {\bm a}_2}{|{\bm a}_1||{\bm a}_2|}
= \frac{g_{12}}{\sqrt{g_{11}} \sqrt{g_{22}}}
\mbox{,} \\
\sin{\Psi} &=&
\frac{|{\bm a}_1 \times {\bm a}_2|}{|{\bm a}_1||{\bm a}_2|}
= \frac{\sqrt{g}}{\sqrt{g_{11}} \sqrt{g_{22}}}
\mbox{,}
\end{eqnarray}
where $g={\rm det}[g_{\mu \nu}]$.
We thus have
\begin{eqnarray}
\cos{(\Psi - \psi)} 
&=& \frac{1}{\sqrt{g_{11} g_{22}}}(g_{12} \cos{\psi}
                               + \sqrt{g} \sin{\psi})
\mbox{,}
\end{eqnarray}
which leads us to the explicit forms of the covariant components
$n_i = {\bm n} \cdot {\bm a}_i$ such as
\begin{eqnarray}
n_1 &=& |{\bm n}||{\bm a}_1| \cos{\psi}
    = \sqrt{g_{11}} \cos{\psi}
\mbox{,} \\
\cr
n_2 &=& |{\bm n}||{\bm a}_2| \cos{(\Psi - \psi)} \cr
 &=& \frac{1}{\sqrt{g_{11}}}(g_{12} \cos{\psi} + \sqrt{g} \sin{\psi})
\mbox{.}
\end{eqnarray}
Finally, the formulas $n^j = n_i g^{ij}$ with $g^{ij} = [g_{ij}]^{-1}$ is 
employed to obtain
\begin{eqnarray}
n^1 &=& \frac{1}{\sqrt{g_{11}}}
        (\cos{\psi} - \frac{g_{12}}{\sqrt{g}} \sin{\psi})
\mbox{,} \\
n^2 &=& \sqrt{\frac{g_{11}}{g}} \sin{\psi}
\mbox{.}
\end{eqnarray}


\end{document}